\documentclass{osa-article}

\journal{osajournal}

\articletype{Research Article}

\begin{document}

\title{360$^\circ$ polarization control of terahertz spintronic emitters using uniaxial FeCo/TbCo$_2$/FeCo trilayers}

\author{Pierre Kolej\'{a}k,\authormark{1,2}  Geoffrey Lezier,\authormark{1} Kamil Postava,\authormark{2} Jean-Fran\c{c}ois Lampin,\authormark{1} Nicolas Tiercelin, \authormark{1} Mathias Vanwolleghem\authormark{1,*}}

\address{\authormark{1}Univ. Lille, CNRS, Centrale Lille, Univ. Polytechnique Hauts-de-France, UMR 8520 - IEMN - Institut d'Electronique de Microélectronique et de Nanotechnologie, F-59000 Lille, France\\
\authormark{2}Technical University of Ostrava, IT4Innovations \& Faculty of Materials Science and Technology, 708 00 Ostrava - Poruba, Czech Republic\\
}
\email{\authormark{*}Corresponding author: mathias.vanwolleghem@univ-lille.fr} 





\begin{abstract}
Polarization control of THz light is of paramount interest for the numerous applications offered in this frequency range. Recent developments in THz spintronic emitters allow for a very efficient broadband emission, and especially unique is their ability of THz polarization switching through magnetization control of the ferromagnetic layer. Here we present an improved scheme to achieve full 360$^\circ$ nearly coherent polarization rotation that does not require multipolar or rotating external magnetic bias nor complex cascaded emitters. By replacing the FM layer of the spintronic emitter with a carefully designed FeCo/TbCo$_2$/FeCo anisotropic heterostructure, we experimentally demonstrate Stoner-Wohlfarth-like coherent rotation of the THz polarization over a full 2$\pi$ azimuth only by a bipolar variation of the strength of the hard axis field, and with only a negligible decrease in the emission efficiency as compared to standard Pt/CoFeB/W inverse spin Hall emitters. THz measurements are in agreement with our model of the non-perfect Stoner-Wohlfarth behaviour. These emitters are well adapted for the implementation of polarimetric characterization not requiring any mechanically rotating polarizing elements. An example is given with the characterization of the birefringence in a quartz plate.
\end{abstract}



\section{Introduction}

THz technology has now entered many societally important applications with exciting demonstrations of nondestructive spectroscopy, chemical sensing, \cite{Dhillon2017, Hindle2018, Jepsen2011, Ellrich2020,nagatsuma2016advances}. Continuous improvements of the performance of the building blocks of a THz analysis chain (sources, modulators, receivers) are bringing many of these demonstrators close-to-market.
Control and flexible manipulation of the polarization state of a THz beam is a functionality in the toolbox that is not straighforwardly realized. Such control would open applications for spectroscopic ellipsometry, coherent control or even polarization multiplexed communications at THz frequencies.\\
THz polarization states can be controlled by Fresnel-rhomb wave plates, wire-grid polarizers, birefringent wave plates or active metamaterial-based devices. While these approaches are robust and can sometimes be actively controlled, they are most often either bulky or narrowband and present important insertion losses \cite{Shan2009, Konishi2020, Grady2013, Markovich2013}. Controlling the emitted polarization state directly upon generation is in this sense a more logical approach. THz polarization shaping has been demonstrated by optical rectification in nonlinear crystals of double- or multiple optical pulses with different polarizations\cite{Amer2005, Shimano2005, Lee2012}, by optical polarization pulse shaping before rectification\cite{Sato2013}, by spatial structuration of the shape of the electrical contacts in a photoconductive switch\cite{Hirota2005} or even by tailoring two-color laser plasma filamentation\cite{Dai2009,Wen2009,You2013}. While all these approaches achieve a high degree of polarization control upon THz generation, their implementation is cumbersome necessitating subtle instrumentation, high power or exhibit limited tunability.\\
Ideally, polarization control should be achieved by an easily accessible degree of freedom of the THz generation mechanism itself. THz emission by the inverse spin Hall effect (ISHE) offers this possibility. Photoinduced ultrafast demagnetization of thin film magnetic heterostructures presenting interfaces between ferromagnetic (FM) transition metals and non-magnetic (NM) 5d metals with strong spin-orbit coupling (SOC), has come to the forefront as a promising new type of THz emitters\cite{seifert2016, Yang2016}. ISHE creates a picosecond spin-to-charge dipole current burst when majority spins are pumped by a femtosecond infrared pulse from the FM into the NM layer. The emitted THz pulse presents characteristics outperforming traditional THz pulsed emitters both with respect to conversion efficiency and bandwidth \cite{Seifert2017, Fulop2020}.\\ 

Moreover, ISHE spintronic terahertz emitters (STE) emit a linear polarization state that is perfectly orthogonal to the FM's magnetization, allowing for straightforward polarization shaping by local control of the magnetization of the FM layer. Rotating linear polarization by 90$^\circ$ degrees has been demonstrated by mechanically rotating the external magnetic saturation field\cite{Yang2016}. An alternative to mechanical rotation of the external magnetic bias has been demonstrated by adding a crossed AC modulated magnetic field to a static magnetic bias\cite{Gueckstock2021}. Spintronic THz polarization modulation over 90$^\circ$ in isotropic CoFeB-based emitters at speeds of 10kHz have been obtained, mainly limited by the reactance of the electromagnet providing the AC magnetic applied field and the need to overcome the coercitivity of the CoFeB. There has been an observation of elliptically polarized THz beams from a CoFeB STE with locally `twisting' in-plane magnetization\cite{KongAOM2019}. 
These could however not fully be predicted by a reproducible, verifiable magnetization distribution.
Applying a quadripolar external magnetic field with opposing polarities on a Ni$_{80}$Fe$_{20}$/Pt bilayer STE induces a well controlled nonuniform magnetization profile with singular polarity points in the center. The measured resulting THz phase front reproduces this exotic polarization profile with very strong field values at the singularity in the centre\cite{Hibberd2019}.
Fully reconfigurable control of circularly polarized THz has been obtained in a two-stage cascaded scheme with orthogonally magnetized STEs that are subtly positioned to obtain the correct phase difference and amplitude equality\cite{ChenAPL2019}. 
These demonstrations of magnetic field leveraged THz polarization control are notable but suffer from a complex implementation or lack of reconfigurability.\\

To overcome these limitations, we propose in this Article, a STE with FM layer that has engineered uniaxial anisotropy. By inducing an in-plane easy axis in a ferromagnetic multilayer with a reasonably low anisotropy field $H_A$, the in-plane magnetization is expected to rotate in a Stoner-Wolhfarth (SW) fashion over a full 2$\pi$ azimuth by applying only a varying external field perpendicular to the anisotropy axis\cite{Stoner1948}. Used as spin pumping FM layer in a STE, such an anisotropic layer greatly eases THz polarization control. To this end, we have developed a Pt/W-based trilayer STE with an anisotropic spin pumping layer based on an intermetallic heterostructure combining rare-earth and transition metals.
We will demonstrate that the azimuth of the emitted linear THz polarization can be coherently rotated over a full circle by simply setting the strength of the hard axis field to the appropriate value. Moreover, the rare-earth intermetallic heterostructure multilayer adds this functionality without sacrificing THz field strength as compared to the record CoFeB-based STE\cite{seifert2016}. An accurate analysis of the magnetic free energy accounting for deviations of the perfect SW monodomain behavior, correctly fits the observed the polarization rotation cycles. This improved scheme for a STE with polarization control does not require any mechanical moving magnets nor special magnetic patterns but a low value simple uniaxial bias. As a result this THz polarization setting is easy to implement in a practical emitter and can extend greatly the potential of this type of emitters for THz spectroscopy and coherent control applications\cite{Kampfrathnatphot2013,Cocker2013,Mosley2017}.

\begin{figure*}[!tb]
    \centering
    \includegraphics[width=\textwidth]{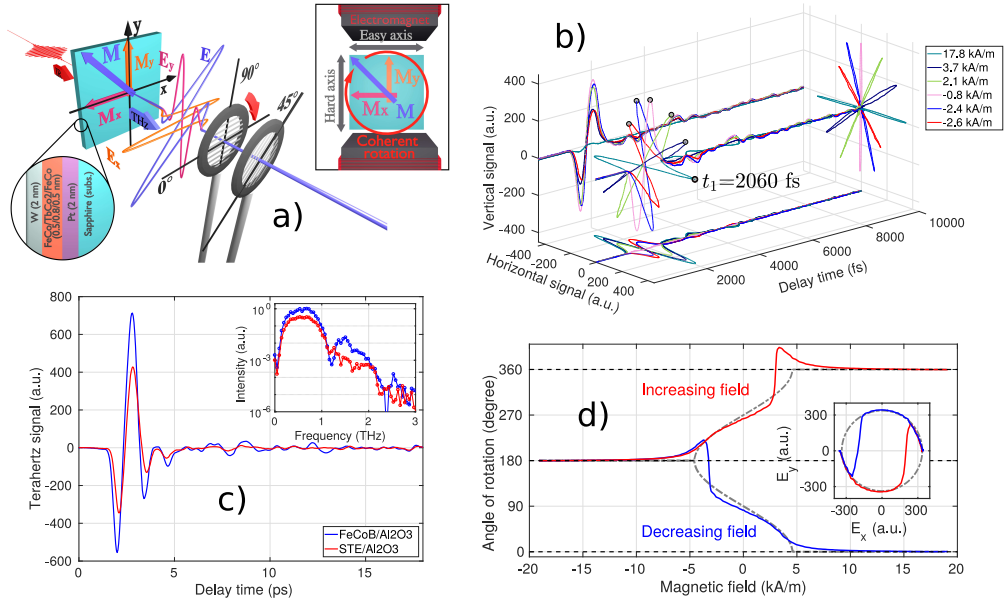}
    \caption{\textbf{Spintronic terahertz polarization rotation:} a) Schematic representation of the spintronic emitter with magnetic uni-axial anisotropy. The STE emitter is excited by an IR pulse from the metallic heterostructure side. This trilayer has an engineered uniaxial anisotropy allowing to control its magnetization direction over a full 360° by only applying a hard axis field as represented in the inset. The polarization of the spin-orbitronic generated terahertz emission $\mathbf{E}(t)$ perpendicular to \textbf{M} therefore rotates accordingly with the applied hard axis field. The polarization components $E_x(t)$ and $E_y(t)$ are detected by 2 wire grid polarizers in series (0/90$^\circ$ switching and 45$^\circ$ fixed). b) 3D representation of the measured time domain traces for the horizontal $E_x(t)$ and vertical components $E_y(t)$ of the transient electric field $\mathbf{E}(t)$  for different values of the externally applied field, showing the rotation of the polarization. c) Comparison of the emitted THz signal between the FeCoB-based reference STE (blue) and the ECML-based emitter (red) in the temporal domain and frequency domain (inset). d) Polarization angle measured as a function of the applied magnetic field. The inset shows the vectorial amplitude of the THZ beam at the fixed delay of 2060fs, illustrating the quasi-lossless full 360° rotation over a hard axis field cycle.
    }
    \label{fig:rotation}
\end{figure*}

\section{THz polarization control using spintronic ferromagnetic heterostructure}

Exchange-coupled multilayer (ECML) heterostructures based on repeated TbCo$_2$/FeCo bilayers grown under a polarizing magnetic field are known to imprint the strong uniaxial anisotropy of the rare-earth alloy on the magnetic multilayer but with a lowered value for the effective anisotropy field $H_{A,\mathrm{eff}}$ and an averaged effective saturation magnetization $M_{s,\mathrm{eff}}$\cite{LeGall2000,BenYoussef2002}. 
Using such an uniaxial ferromagnetic multilayer as spin pumping layer in an ISHE STE is expected to lead to a linear THz polarization that will emulate the hysteresis of the multilayer.
In particular, if perfect monodomain uniaxial anisotropy is achieved, it is known from magnetism that according to the Stoner-Wohlfarth model the magnetization of the ECML will uniformly rotate over 180$^\circ$ by varying the applied field only along the hard axis in the range of the anisotropy field strength $[-H_{A,\mathrm{eff}},+H_{A,\mathrm{eff}}]$. Specifically we have grown a W(2nm)/CoFe(5\AA)/TbCo$_2$(8\AA)/CoFe(5\AA)/Pt(2nm) stack on a c-cut sapphire substrate. During the growth a magnetic induction $\mu_0 H_d\approx$ 0.1T is applied in the plane of the substrate to induce an easy axis direction in the anisotropic layer (see Supplemental Information).
 The 5d metals Pt and W provide the ISHE with opposite signs for $\theta_{SH}$\cite{Hoffmann2013}. Placed on opposite sides of the ferromagnetic spin pumping layer, they will therefore convert incoming spin currents into charge currents of the same sense, as follows directly from $\vec{J}_C = \theta_{SH}\frac{2e}{\hslash}\vec{J}_S\times\vec{\sigma}$, where $\vec{J}_C$, $\vec{J}_S$ and $\vec{\sigma}$ are respectively the charge current density, the injected spin current density and the spin polarization direction\cite{Hoffmann2013}. Figure~\ref{fig:rotation}a illustrates the proposed principle of the polarization controlled spintronic emitter. The sample is mounted between the pole shoes of an electromagnet providing a uniform external biasing magnetic field H$_\mathrm{ext,max}=40\tfrac{\mathrm{kA}}{\mathrm{m}}$ (i.e. B$_\mathrm{ext,max}=50$mT). 
Its operation is verified on a customized time-domain spectroscopy setup (see Supplemental Information). The emitter was excited by femtosecond infrared pulses from a Ti:sapphire oscillator (100fs pulses, centre wavelength 820nm, repetition rate 80MHz, average power 185mW) that are normally incident on the metallic ECML stack. The emitted THz pulses were observed from the substrate side in a transmission geometry. In this way, echoes in the time domain transients are strongly suppressed since the metallic stack behaves as a good antireflective coating for THz frequencies \cite{carli1977reflectivity,kroll2007metallic}. The transient electric field emitted by the magnetically anisotropic ECML was coherently sampled using a photoconductive low-temperature grown GaAs dipole antenna. A couple of wideband wiregrid polarizers allow to select between the orthogonal polarization components of the emitted THz wave (see Supplemental information). Throughout this work, the convention is used that the (horizontal) x-axis is aligned with the induced anisotropy of the ECML. The x- and y-axes in this work can thus also be read as easy-axis (EA) and hard-axis (HA) directions.

Figure~\ref{fig:rotation}b plots the measured horizontal (x) and vertical (y) components, $E_x(t)$ and $E_y(t)$, of the transient electric field $\mathbf{E}(t)$  for different values of the external applied field $H_{\mathrm{ext},y}$. 
When projecting these traces as vectorial parametric curves on a 2D plane as is done in this figure, the obtained polarization cycles illustrate the main points of the suggested operation principle. Firstly, as expected from theory, the horizontal and vertical components are in phase. At all applied field strengths, $E_x(t)$ and $E_y(t)$ reach their zeroes and extrema at the same time delays. The generated charge current pulse acts as a dipole emitter exactly perpendicularly polarized to the polarity of the spin current without any phase difference. The emitted polarization is thus perfectly linear. The very small ellipticity observed at $3.8\tfrac{\mathrm{kA}}{\mathrm{m}}$ is the result of instability in the voltage source controlling the electromagnet combined with heating problems in the coils. As a result, the applied field (and thus the sample's magnetization) was not stable over the time of the measurement. Note also that there is no possible optical THz anisotropy induced by the sapphire substrate since it is c-cut and the emitter is observed along the z-axis. Secondly, as the applied external magnetic field $H_{\mathrm{ext},y}$ is decreased the polarization plane starts gradually rotating from being perfectly aligned with the x-axis towards the y-axis (as $M_y$ rotates towards $M_x$).
In order to visualize this rotation more clearly, we have plotted in Fig.~\ref{fig:rotation}d as a function of $H_{\mathrm{ext},y}$ the THz polarization azimuth $\varphi$ with respect to the x-axis obtained by $\varphi=\left.\arctan\left(\frac{E_y}{E_x}\right)\right|_{t=t_1}$, evaluated at a chosen fixed time delay (for convenience an extremum of the transient signal, $t_1=2060\mathrm{fs}$). 
In doing so, it is assumed that the dynamics of the spin-to-charge conversion are up to first order unaffected by the polarization of the spin current. In other words there is no anisotropy in the generated dipole current burst in the heavy nonmagnetic metal layers and the spin Hall angle is a scalar isotropic constant\cite{neumann2016temperature,hao2015giant}. The temporal traces of Fig.~\ref{fig:rotation}b 
indicate that indeed all $\mathbf{E}(t)-$traces reach their extrema at equal time delay, independent of the applied $H_{\mathrm{ext},y}$. The measured hysteresis exhibited by the polarization azimuth as a function of hard axis field proves that the proposed emitter allows to rotate the linear THz polarization over a full circle by only a modest variation of an uniaxially applied magnetic field within a range of just $\pm10\tfrac{\mathrm{kA}}{\mathrm{m}}$ $(\mu_0 H \approx \pm 12\mathrm{mT})$. The inset polar plot of the vectorial $\mathbf{E}$-field amplitude proves that this hysteretic rotation is quasi-perfect except near the switching field. Moreover inspecting the behaviour in Fig.~\ref{fig:rotation}d, $\varphi$ seems to follow a $\cos\varphi\propto \frac{H_{\mathrm{ext},y}}{H_A}$ with $H_A\approx 5\tfrac{\mathrm{kA}}{\mathrm{m}}$, as indicated by the dashed-dotted cosine curve. Thus, $E_x = |\mathbf{E}|\cos\varphi$ depends linearly on the applied hard axis field. According to the physics of the ISHE, this implies that $\sigma_y$, the HA projection of the polarization of the injected majority spin current pulse, likewise depends linearly on the HA field $H_{\mathrm{ext},y}$. Such a behaviour is reminiscent of a Stoner-Wohlfarth rotation of the magnetization of the FM stack. The measurements of Figs.~\ref{fig:rotation} 
suggest that this mechanism is transposed to the emitted THz radiation. We will show in the next section that the THz polarization is indeed an exact "carbon copy" of the magnetostatic behaviour of the FM ECML stack. Polarization control in THz ISHE emitters can therefore be engineered by controlling the magnetic anisotropy of the spin pumping layer. The deviation of the perfect cosine-behaviour close to field strengths of about $\pm 2.5 \tfrac{\mathrm{kA}}{\mathrm{m}}$ is due to a combination of non-ideal monodomain behaviour and rapid switching of the ECML stack.  \cite{explanation_tilt}
This will be further discussed in Section~\ref{sec:SW_model}

\begin{figure*}[!b]
    \centering
    \includegraphics[width=\textwidth]{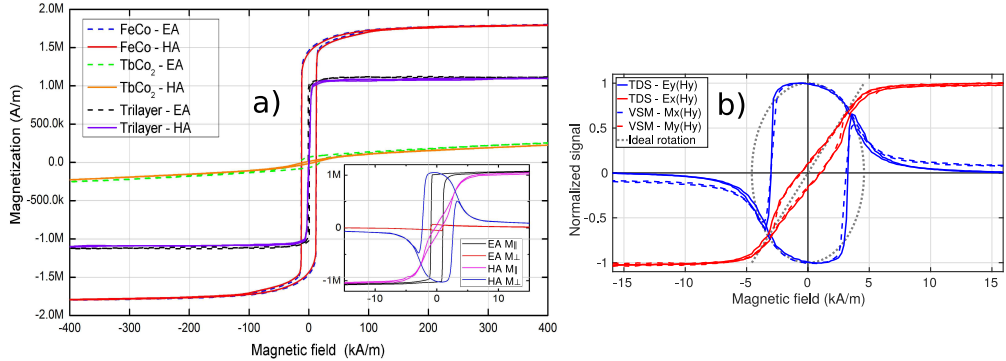} 
    \caption{
     \textbf{Terahertz magnetostatic hysteresis of the uniaxial FeCo/TbCo$_2$/FeCo heterostructure:} a) VSM measured magnetization characteristics for single layers of FeCo, TbCo$_2$, and a FeCo(5\AA)/TbCo$_2$(8\AA)/FeCo(5\AA), all sputter deposited under a polarizing magnetic field inducing an easy axis (EA). The main graph shows the M$_\parallel$-component along the applied field direction, oriented for all layers either along the induced EA or orthogonal to it (HA). Exchange coupling in the trilayer results in an engineered uniaxial anisotropy with averaged $M_s$ and lowered $H_A$. The inset zooms in on the trilayer's vectorial hysteresis, evidencing its well defined uni-axial anisotropy. For HA fields the $M_\parallel$ (pink) and $M_\perp$ (blue) hysteresis exhibit near perfect SW rotation with an anisotropy field of 5kA/m. b) Comparison between magnetization hysteresis loops measured using VSM (dashed lines) and E-field THz-TDS measurements (full lines) at the maximal peak position. The grey dotted line illustrates the ideal circular hysteresis. The magnetostatic properties of the spin pumping layer are perfectly recovered in the polarization behaviour of the spintronic terahertz generation.
     }
    \label{fig:3dloop}
\end{figure*} 
In any case, the measured one-to-one correspondence between polarization azimuth and applied field strength allows to set the polarization angle at will, not needing any mechanically rotating parts nor complex cascading of dephased THz emitters. Spurious crosspolarization is not observed or at least better than the 30dB rejection rate of the wiregrid polarizers used \cite{wiregrid}.
The limited field strengths needed to cover $2\pi-$rotations open possibilities to implement non-quasistatic rotations of the THz polarization. This finds applications in THz polarimetry and ellipsometry, or as a THz emitter with built-in fast polarization modulation using ferromagnetic resonance (FMR) techniques\cite{KlimovIgnatov2010}. An example of its use in polarimetry will be given in Section~\ref{sec:application}.

\section{Uniaxial anisotropic inverse spin Hall emitter: performance studies}
\label{sec:TDS_measurements}

With the THz polarization rotation established, we turn our attention to the building blocks allowing this mechanism and assess their performance. 
The impact of the use of an exchange coupled anisotropic multilayer as spin pumping layer, instead of a single layer ferromagnetic isotropic transition metal alloy, needs to be evaluated. Within the family of intrinsic ISHE THz emitters those based on (W/FM/Pt)-trilayers are known to achieve the strongest spin-to-charge current conversion (and hence THz power)~\cite{Dang2020}. In Fig.~\ref{fig:rotation}c the anisotropic ECML-based emitter is compared to the reference record ISHE trilayer W(2nm)/Co$_{20}$Fe$_{60}$B$_{20}$(1.8nm)/Pt(2nm) emitter~\cite{Seifert2017} by measuring its transient electric field on the TDS setup under identical experimental conditions. The reference CoFeB emitter was grown on the same type double-side polished c-cut sapphire substrate. The 5d non-magnetic SOC layers are the same in both emitters. 
The total thickness of the FM stack in the rare-earth based emitter is the same as that of the optimized isotropic CoFeB emitter. Both emitters are magnetically saturated along their easy axis ($H_\mathrm{ext}=\tfrac{200}{4\pi}\tfrac{\mathrm{kA}}{\mathrm{m}}$). They were excited from the metal side in order to suppress THz echoes in the TDS measurements~\cite{torosyan2018}. 
As the THz emission is governed by the spin-to-charge conversion in the 5d SOC metals and since both emitters use the same ones and in the same stacking order, both measured E-field transients present the same polarity and identical temporal dynamics (see Fig.~\ref{fig:rotation}c).\\
Remarkably, the TbCo$_2$/FeCo-based emitter achieves a peak-to-peak THz signal that is as high as 60\% of that of the reference CoFeB trilayer. The Fourier transform (plotted as inset in Fig.\ref{fig:rotation}c) confirms this observation. It is known that THz pulses generated by the ISHE in 5d metals are quasi Fourier-limited in view of the very short electron lifetimes in the metal layers~\cite{Dang2020}.  The bandwidth of the spectra is therefore dominated by the response time of the Auston switch of the TDS setup (see Supplemental Information) and not by the dynamics of the ISHE. Nevertheless, the observed bandwidth of about 3THz is close to the minimum time-bandwidth product of the 100fs pulses in the custom TDS setup ($\Delta f \approx 4$THz). 

The measured spectral width of the ECML-based emitter is unchanged with respect to the reference CoFeB-based emitter, underlining again that dynamics is governed by the 5d metal layers. When assessing the observed THz efficiency of the ECML emitter, it was pointed out by Seifert in~\cite{seifert2016} that ISHE THz generation presents an optimum as a function of total metal thickness. The other parameters governing this optimum formula are all intrinsic material properties (spin Hall angle and THz refractive indices or conductivities). This was refined by Torosyan~\cite{torosyan2018} to take into account the individual FM and NM layer thicknesses and their conductivities. Making a reasonable assumption that the conductivity of the TbCo$_2$/FeCo multilayer is not radically different from that of the amorphous CoFeB alloy and also ignoring self-ISHE in the spin pumping layer, and all the rest being equal in both emitters, the difference in THz efficiency is thus entirely attributed to a less efficient spin current generation and injection for the ECML emitter. This can have multiple origins. An impact of the two extra interfaces in the ECML in the form of spin scattering losses in the superdiffusive spin current is likely to occur. But the dominant factor is believed to be the lower net saturation magnetization of the ECML, leading to a lower maximal absolute spin polarization injection in the 5d metals. 

\begin{figure*}[!b]
	\centering 
	  \includegraphics[width=\textwidth]{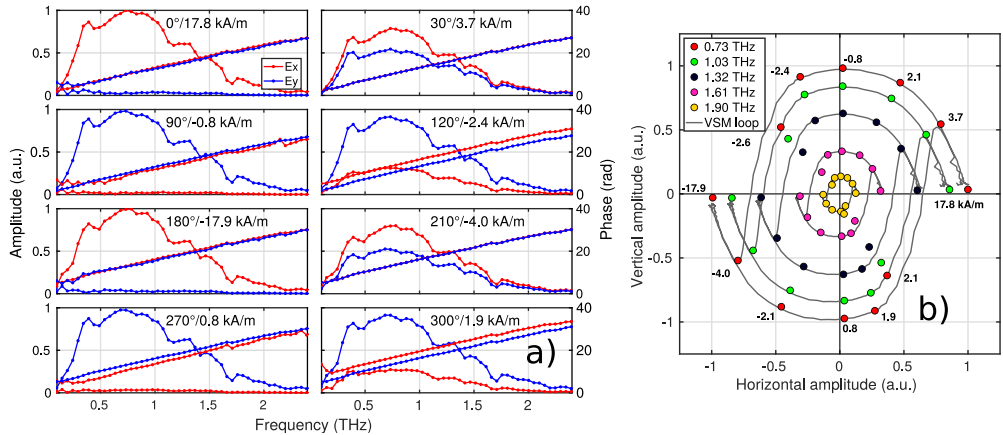} 
	\caption{\textbf{Analysis of the terahertz polarization rotation in Fourier space:} a) evolution of normalized spectral amplitudes and extracted phases of $E_x$ (red) and $E_y$ (blue) components during a full hard axis cycle (for 8 rotation angles/magnetic fields distributed over a full circle). Note that the phase extraction for near-zero components is saddled by numerical noise. Still it is clear that both components are always perfectly in-phase or anti-phase depending on the quadrant angle of the polarization, underlining the perfect linearity of the polarization. b) Cartesian paramteric plot of the vectorial Fourier amplitude $\mathbf{E}(\omega)=\mathbf{1}_x|E_x(\omega)|+\mathbf{1}_y|E_y(\omega)|e^{j\Delta\phi(\omega)}$ for selected frequencies up to 2THz. As $\Delta\phi = 0,\pi$ there is no ellipticity and $\operatorname{Im}\mathbf{E}(\omega)=0$. The Stoner-Wohlfarth rotation is recovered trhoughout the spectral content of the pulse. Homothetically scaled VSM hysteresis curves are added as guide to the eye.}
	\label{fig:fourierphase}
\end{figure*}
In order to corroborate this assumption we have studied the static magnetization properties of the stack using an ADE EV9 vectorial vibrating sample magnetometer (VSM) and compared it to those of its constituents. For that purpose, separate samples of CoFe and TbCo$_2$ have been grown under the same polarizing field $H_d$ using the same method as the ECML. Figure~\ref{fig:3dloop}a plots the magnetization component for all three systems (FeCo, TbCo$_2$ and FeCo/TbCo$_2$/FeCo) measured along the direction of the magnetic field, when it is applied either along the induced easy axis (EA) or perpendicular to it (HA). These data demonstrate clearly how the coupling strongly lowers the effective anisotropy field of the rare earth alloy, and averages the $M_s$ of the complete stack. The $M_s$ of the ECML is indeed lowered to about 60\% of that of the amorphous FeCo alloy explaining the lower spin polarization in the anisotropic emitter. The exchange coupling between thin films of intermetallic rare earth and transition metal alloys is thus a very efficient mechanism to introduce anisotropic control in ISHE emitters without sacrificing significantly the spin injection mechanism.
The zoom on the ECML hysteresis in the inset of Fig.~\ref{fig:3dloop}a makes its uniaxial anisotropy more apparent. An anisotropy field reduced to approximately $5\tfrac{\mathrm{kA}}{\mathrm{m}}$ is observed while maintaining a saturation magnetization close to that of the isotropic CoFe system ($M_s\approx \tfrac{1\mathrm{MA}}{\mathrm{m}}\approx 0.6 M_{s,\mathrm{CoFe}}$). By using vectorial VSM probes, this uniaxial anisotropy is also observed in the near perfect SW-like rotation of the magnetization in the plane of the ECML. Varying the applied magnetic field perpendicular to the induced easy axis, the magnetization component along this direction ($M_y$) varies almost perfectly linearly between $\pm H_A$, while the orthogonal component ($M_x$) simultaneously follows nearly a $M_x(H_\mathrm{ext})=\sqrt{1-M_y^2(H_\mathrm{ext})}$-law for $H_\mathrm{ext} \leq H_A$. This is signatory SW-behaviour \cite{Stoner1948}.
The observed deviations are the combined effect of a small (intentional) misalignment (see \cite{explanation_tilt}) between the applied field and the anisotropy axis and the appearance of multidomains during magnetization reversal as will be seen in Sec.~\ref{sec:SW_model}.

The physics of the ISHE imply that the emitted THz polarization should perfectly emulate this magnetostatic Stoner-Wohlfarth rotation. While it cannot be straightforwardly deduced from the temporal vectorial polarization traces as a function of applied hard axis field in Fig.~\ref{fig:rotation}b, it was hinted at by the behaviour of the azimuth of the polarization (see the inset in Fig.~\ref{fig:rotation}d). To make this fully visible, we compare in Fig.~\ref{fig:3dloop}b normalized hard axis hysteresis cycles obtained by vectorial VSM magnetometry (dashed lines) of the in-plane $M_x$ and $M_y$ components (refer to the inset in Fig.~\ref{fig:3dloop}a) with the normalized $E_y(t_1)$ and $E_x(t_1)$ THz polarization components (full lines) measured via TDS (at the same time delay $t_1 = 2060$fs in Fig.~\ref{fig:rotation}b). Clearly the TDS vectorial polarization cycles reproduce exactly VSM magnetometry. The difference in asymptotic behaviour between $E_x$ (TDS) and $M_y$ (VSM) are not related to ISHE physics but are an artefact of the VSM measurements. VSM is a volumetric method and therefore the slight paramagnetic contribution of the substrate is also picked up. This one-to-one correspondence between VSM magnetometry and TDS THz polarimetry in ISHE emitters is a striking result that reconfirms that the THz polarization rotation occurs without signal loss. Superposed on the TDS and VSM hysteresis cycles, the dotted curves illustrate ideal SW rotation with an anisotropy field close to 5$\mathrm{\tfrac{kA}{m}}$. 
As indicated previously, the magnetization flip during the rotation prevents a perfect Stoner-Wohlfarth single domain rotation. Indeed, in the narrow window where the jump occurs, the polarization is extremely sensitive to the external field. Still, a 360° rotation is observed.

Up till now the discussion has been limited to the time domain, analysing the polarization behaviour at a fixed point in the transient signal. Even though the temporal traces in Fig.~\ref{fig:rotation}b seem to suggest that both THz field components always remain in phase, this can only be correctly assessed by Fourier analysis. Moreover the spectral behaviour of the polarization is crucial for spectroscopic and ellipsometric applications. Figures~\ref{fig:fourierphase}a plot the amplitude and the phase of the Fourier transform of the measured field components $e_{x,y}(t)$ at 8 different values of $H_{\mathrm{ext},y}$ leading to polarization angles distributed over the four quadrants and aligned along the cartesian axes, as extracted from Fig.~\ref{fig:rotation}d. 
The amplitudes have been normalized to the maximum of the spectrum when the sample is saturated along the hard axis. The spectra are plotted in a linear scale in order to make the linearity of $E_x$ (red spectra) with $\tfrac{H_{\mathrm{ext},y}}{H_A}\: (H_A\approx 5\mathrm{\tfrac{kA}{m}})$ more evident. It is already qualitatively clear how the field spectra transform almost unperturbed over the (xy)-cartesian plane as a function of HA field.
 In order to make this more obvious, Fig.~\ref{fig:fourierphase}b plots a cartesian projection of the evolution of the spectral amplitudes from Figs~\ref{fig:fourierphase}a in a cartesian plane for frequencies every 0.3THz between the spectrum peak (0.73THz) and 2THz. Note that this parametric plot takes into account the phase difference $\Delta\phi = \arg(E_y)-\arg(E_x)$ by plotting the real part of the vector $\mathbf{E}(\omega)=\mathbf{1}_x|E_x(\omega)|+\mathbf{1}_y|E_y(\omega)|e^{j\Delta\phi(\omega)}$. As expected, the observed Stoner-Wohlfarth rotation in the time domain (see Fig.~\ref{fig:rotation}d) is recovered here in the spectral content of the pulses. While the measured vectorial spectral hysteresis is plotted at only a few selected field values, a homothetically scaled version of the vectorial VSM cycle acts as a guide to the eye proving the presence of SW rotation over the whole spectral content of the pulse. In order to make this argument even stronger, the Fourier spectra in Fig.~\ref{fig:fourierphase}a also plot the phase of both components. First of all, the phase behaviour is featureless (except for the natural linear increase due to the air propagation between emitter and detector) which correctly implies that the ISHE pulses are Fourier transform limited due to the extreme bandwidth of the ISHE in SOC metals. Secondly, both components exhibit a perfect in- or antiphase behaviour over the entire spectrum. Except for the cases where the polarization is purely x- or y-oriented (when one of either components is close to the noise floor), the phase difference is flat and nearly perfectly 180$^\circ$ or 0$^\circ$. This also proves the basic assumption of ISHE physics, namely that the  
spin-to-charge current burst behaves as a point-like dipole source without any feasible phase shift during the magnetization rotation. 
All of the above proves that the linearity of the polarization is stable over the entire spectrum of the pulse and during the full coherent $2\pi-$rotation.

\section{Extended Stoner-Wohlfarth incoherent rotation fitting for ECML spintronic emitters}\label{sec:SW_model}

\begin{figure*}[h!]
    \centering
    \includegraphics[width=\textwidth]{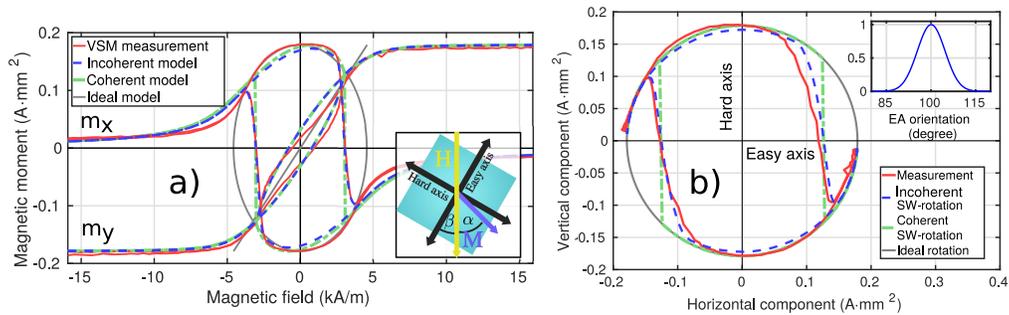}
    \caption{\textbf{Hysteresis fitting:} comparison of incoherent (blue dashed) and coherent (green dot-dashed) fitted models with VSM measurement (red line) for a) both $m_x$ and $m_y$ and b) the magnetic moment rotation (grey line represents the ideal circular rotation). The geometrical meaning of the angles $\alpha$ and $\beta$ is schematized as the inset of a). To illustrate the phenomenological extended multidomain incoherent SW model,  the inset of b) shows the fitted Gaussian distribution of the easy axis orientation $\beta_i$ of the domains.}
    \label{fig:model_comp}
\end{figure*}

Having established the uniaxial field control of the THz polarization over a full $360^\circ$, we now demonstrate how the hysteresis loop can be perfectly modeled. This is of practical importance. In view of the demonstrated perfect orthogonality between the emitted linear THz polarization and the magnetization of the multilayer, such a model would allow a one-to-one relationship between HA field and polarization azimuth, that can be used for instance in polarimetric applications.
The starting point is the assumption that the ECML rotates according to the Stoner-Wohlfarth model as a single domain with an uniaxial in-plane magnetic anisotropy axis\cite{Stoner1948}. This is motivated by the observations above. In the following we will elaborate an extended multidomain SW model that correctly fits the observed slight deviations. The equilibrium position of the magnetization of a given domain is described by the minimization of the magnetic energy $E_M$,
\begin{equation}
E_M(\alpha,\beta,\mathbf{H})=-\mu_0 |\mathbf{H}| m_S \cos(\alpha )+K_u \, V \, \sin^2(\alpha-\beta),
\label{eq:SW}
\end{equation}
where $\mu_0$ is the vacuum permeability and the material domain parameters $m_S$, $K_u$, and $V$ are the saturation magnetic moment, the uniaxial anisotropy constant, and the domain volume respectively. The angles $\alpha$ and $\beta$ denote respectively the orientation of the magnetization and of the induced easy axis with respect to the external magnetic field $\mathbf{H}$ as illustrated in the inset of Fig.\ref{fig:model_comp}. The first term of Eq.~\ref{eq:SW} is the standard Zeeman field energy describing the potential energy in an external magnetic field. The second term describes the uniaxial anisotropic energy, assuming the coupled multilayers behave as an effective system with a magneto-crystalline energy having a 180-degree rotation symmetry. This symmetry originates in the induced magnetic anisotropy by interlayer coupling, which implies an analogy with the magnetic anisotropy of tetragonal lattice systems.\cite{cullity2011introduction}.

If the coupled multilayers behave as a perfect single magnetic domain, this Stoner-Wohlfarth model predicts the saturation magnetization to uniformly rotate as function of applied field with an abrupt switch when the Zeeman energy overcomes the anisotropy energy. This sudden switch depends on the angle between the anisotropy axis and the magnetic field. It becomes less pronounced as the anisotropy axis approaches orthogonality with the applied field. In the latter limiting case the magnetization perfectly coherently rotates over full $2\pi$ angle. In practice the magnetization reversal is the result of a complex interplay of domain nucleation, domain wall motion, curling or buckling and other reversal processes that can only be correctly modeled by micromagnetic simulations \cite{vaz2008magnetism,arrott1991phenomenology}. It is beyond the scope of this work to account for these processes. Instead we propose in the following a pragmatic phenomenological modeling approach that allows to fit with reasonable accuracy the observed magnetization reversal. The VSM-measured magnetization switching presents a somewhat rounded and smoothed out behaviour. Especially close to the critical field values the $m_x$ hysteresis curves of Fig.~\ref{fig:model_comp}(a) (full red line) can be seen as the averaged switching of an ensemble of polycrystalline domains with distributed easy axes around $\beta=\pi/2$.

Taking this phenomenological approach to describe the magnetization reversal, the hysteresis loops are approached by a incoherent sum of magnetic moments $m_i$ of multiple domains with a normal distribution of their easy axes $\beta_i$ about an average axis $\beta_c$. 
\begin{equation}
g_i(\beta_i)= \mathrm{exp}\left [-\frac{(\beta_i-\beta_c)^2}{G^2}\right  ], 
\end{equation}  \\   
where $G$ and $\beta_c$ are the Gaussian width and the central angle of easy axes of domains, respectively. Each domain is assumed to contribute the same saturation magnetic moment $m_s$, just as it is assumed that the domains extend over the thickness of the exchange coupled stack.
The jump accompanying the magnetization rotation when $\beta\neq\pi/2$, is now smoothed out by the creation of multi-domains and slightly different easy axis orientation of magnetic domains in the whole plane of the ferromagnetic stack. 
This results in the following fitting model for the x,y-component of the effective magnetic moment $\mathbf{m}_\mathrm{eff}$ of the ferromagnetic heterostructure as a non-coherent summation of distributed magnetic moments $\mathbf{m}_i$.

\begin{equation}
m_{\mathrm{eff}}^{\mathrm{x,y}}=\frac{\sum\limits_{i=1}^{N} g_i(\beta_i) 	\, m_i^{\mathrm{x,y}}(\alpha_i, \beta_i)}{\sum\limits_{i=1}^{N} g_i(\beta_i)},
\label{eq:MD}
\end{equation}
where 

\begin{equation}
m_i^\mathrm{x}=m_S \, \mathrm{cos}(\alpha_i) \qquad m_i^\mathrm{y}=m_S \, \mathrm{sin}(\alpha_i)
\end{equation}\\ 
for the external magnetic field $\mathbf{H}$ aligned vertically. The fitting consists of a two-step procedure where for a chosen normal distribution (with parameters $(\beta_c, G)$), N domains are uniformly sampled with easy axis angles $\beta_i$. The magnetization orientation of the $i-$th domain ($\alpha_i$) is then obtained minimization of the magnetic energy $E(\alpha_i,\beta_i,\mathbf{H})$ described by the Stoner-Wohlfarth rotation (\eqref{eq:SW}) for each domain separately and for all field values of the cycle. The noncoherent sum \eqref{eq:MD} is then compared to the VSM data over the complete hysteresis cycle. This procedure is repeated until convergence (see Supplemental Information).\\

 The vectorial VSM hysteresis loops (solid red lines in Fig.\ref{fig:model_comp}), measured along the hard axis, are fitted both to a single domain (\eqref{eq:SW}) and the above multidomain SW model (\eqref{eq:MD}), with $M_s$ and $K_u$ as fitting parameters for the multilayer treated as a single effective uniaxial anisotropic layer. In the MD model the phenomenological statistical parameters $\beta_c$ and $G$ are the two additional fitting parameters. Fig.~\ref{fig:model_comp} plots the results of this coherent (green dot-dashed) and incoherent (blue dashed) Stoner-Wohlfarth rotation. The MD model clearly fits the magnetization reversal better. Combining the horizontal and vertical hysteresis loops and their fits in an in-plane parametric rotation plot in Fig.\ref{fig:model_comp}(b) illustrates how the typical abrupt switching of monodomain SW rotation fails to correctly grasp the measured reversal. While there remain some residual differences between the incoherent model and the VSM data, the ease of implementation of a multidomain model with normally distributed anisotropy axes combined with its reasonable fitting proves that this is a powerful tool to obtain effective material parameters even without any knowledge about the interlayer coupling process~\cite{kools1996exchange}. Obviously only an advanced micro-magnetic approach (domain walls, interlayer coupling, defects, temperature influence, etc.) would be needed to describe in detail the magnetic anisotropic rotation. This is both beyond the scope of this work and is of no practical importance for the targeted application. 

The inset in Fig.~\ref{fig:model_comp}(b) shows the normalized easy axis distribution corresponding to the fitted magnetization rotation. The obtained central angle $\beta_c=100.06^\circ$ is off the ideal circular rotation (grey line) by 10$^\circ$, which was arbitrarily chosen to demonstrate the magnetization switching and the effectiveness of the model\cite{explanation_tilt}. 
The relatively important smoothing of the experimental hysteresis curves is confirmed by the fitted value for the Gaussian width of the normal distribution of $G=6.598^\circ$. Finally, an effective saturation magnetic moment $m_S=0.179$~A$\cdot$mm$^2$ and anisotropy energy of $K_u V=514.2$~pJ are extracted. This implies a saturation magnetization $M_s=m_s/V\approx994.5$~kA/m and an anisotropic constant $K_u\approx2.857$~kJ/m$^3$ given the ferromagnetic ECML thickness of 1.8~nm and a sample surface of $\approx10\times10$~mm$^2$. These values are in agreement with was typical found in literature for Co-based tetragonal systems~\cite{boyd1960magnetic,cullity2011introduction, heinrich2006ultrathin}.

\section{Application example and perspectives}\label{sec:application}

\begin{figure*}[!tbp]
    \centering
    \includegraphics[width=\textwidth]{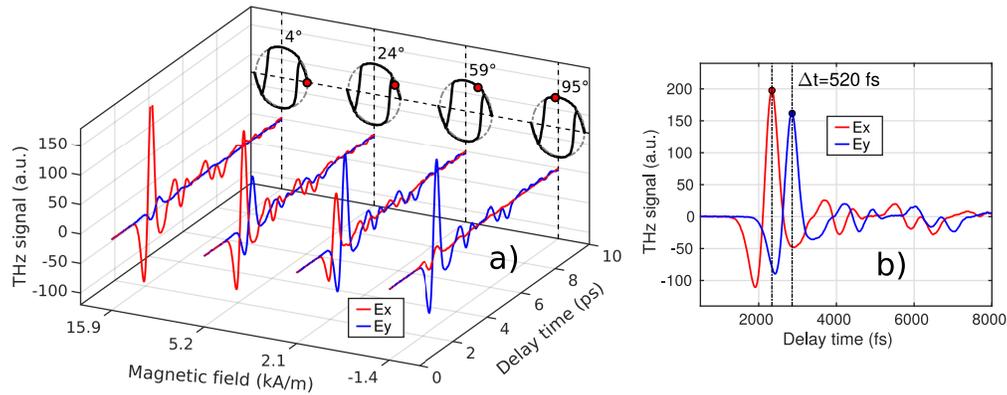}
    \caption{
    \textbf{Terahertz birefringence characterisation of quartz:} the uniaxial magnetic field control of the polarization state of the spintronic emitter is exploited to probe the ordinary and extra-ordinary wave delay of quartz without any rotating polarizing elements. a) Evolution of the detected transient trace (decomposed into its cartesian components) as a function of hard axis field on the emitter. The correponding polarization angle is obtained from the model in Sec.~\ref{sec:SW_model} and indicated by a red mark on the amplitude rotation profile (black line). The gradual transition of the character of the transient from ordinary to extraordinary delay is strikingly visible. b) The extracted phase shift between the fast and slow axis of orthogonal polarizations obtained by simply switching the hard axis field between saturation and near-zero.
    }
    \label{fig:quartz}
\end{figure*}

The easy control of the linear THz polarization by a simple variation of an uniaxial external field finds several important applications in THz technology. One straightforward example is the characterisation of the birefringent properties of a material without the need for rotating polarizing optical elements. Quartz is a well known uniaxial birefringent material used in a very wide frequency range for waveplate applications. In the THz band quartz moreover presents excellent low absorption\cite{Grischkowsky1990, Davies2018}. Typical THz characterization of quartz' birefringence consists of mechanically aligning its optic axis with respect to the polarization of the probing THz beam. This is prone to misalignments and related inaccuracies. The here presented spintronic emitter permits a polarimetric characterization of the birefringent properties by a simple sweep of the applied magnetic field bias on the THz emitter, given the near perfect 360$^\circ$ rotation of the linear THz polarization over this sweep. As such both the ordinary and extraordinary rays in a quartz plate can be probed without any mechanical rotation of either the sample or a polarizing optical element. Given moreover the broadband character of the spintronic emitter, this allows for a wideband THz chacterization of the birefringence. 

We have characterized a 3mm thick quartz plate with its optical axis approximately aligned with the magnetic anisotropy axis of the spintronic emitter (throughout this paper chosen as the x-axis). Fig.\ref{fig:quartz} illustrates as a function of applied magnetic field the evolution of the THz transients from the spintronic emitter when passing through the quartz plate. To facilitate the analysis and to better illustrate the potential of the emitter, the measured transient has been decomposed into its constituting vectorial components (using a switching$\pm 90^\circ$ and fixed $45^\circ$ WGP before the Auston switch dipole as described in Supplemental Information). This can also be achieved without two polarizer settings using electrooptical sampling\cite{planken2001}. Using the fitted model of Section\ref{sec:SW_model} the polarization state of the THz pulse incident on the quartz plate is known for each field setting. This is indicated in Fig.~\ref{fig:quartz} by the point on the hysteresis curve per measured transient. As the HA field $H_{\mathrm{ext},y}$ is gradually reduced from near saturation (15.9kA/m) towards near-zero (-1.4 kA/m) the magnetization evolves from near HA- to near EA-aligned, or the polarization from near x-oriented to near y-oriented. The linear THz polarization thus varies from being almost aligned with quartz's optic axis to being almost perpendicular to it. The measured transient pulse on the TDS setup therefore presents a group delay evolving from a pure ordinary fast wave to a pure extraordinary slow wave while passing through a regime where both delays are hybridly mixed. This is clearly visible in the four decomposed transients in Fig.~\ref{fig:quartz}a. Near HA saturation of the emitter (i.e. polarization parallel to the optic axis of quartz), the transient measured on the Auston switch exhibits principally the delay characteristics of the ordinary (x-polarized) wave in quartz. The decomposition nicely shows that in agreement with the point on the hysteresis curve, a small contribution of the extraordinary wave is expected due to a small magnetization component along the EA (or thus a small y-polarized wave). This small $e_y$-transient (blue trace) at 15.9kA/m appears as a temporally shifted pulse with the same phase as the principal $e_x$-transient. For the subsequent field values, Fig.~\ref{fig:quartz}a illustrates the gradual increase of the $e_x$-transient in agreement with the evolution of the emitted THz polarization. It is striking to see how the pulse evolves nearly unaffected to a temporally shifted copy. One also notices how at the $95^\circ$ polarization the x-transient pulse has taken on the opposite phase in accordance with the polarization entering the second quadrant!

Fig.~\ref{fig:quartz}b focuses on this temporal delay for the limiting cases of nearly pure ordinary and extraordinary ray excitation. It can be seen that the emitted THz pulse is uniformly shifted as expected. The observed group delay shift $\Delta t_g=520\mathrm{fs}$ should correspond to a difference of group index of $\Delta(n_g) = \Delta t_g \times c/L$. With $L=3000\mu$m and $c\approx300\mu$m/ps this gives $\Delta(n_g)\approx 0.052 = \Delta n_{e,o} + \omega\Delta(\tfrac{\partial n_{e,o}}{\partial \omega}) $. As up to first order the dispersion of quartz's ordinary and extra-ordinary index is equal, the group index difference equals quartz's THz birefringence $\Delta n_{e,o}$. The obtained value of 0.052 reproduces the birefringence of quartz close to 0.75THz (the pulse's central frequency) as reported in literature~\cite{Grischkowsky1990,castro2009extraction}.

The goal of this experiment was mainly to illustrate the potential of the emitter allowing fast and efficient characterization of anisotropic materials without needing any polarizing elements or rotation of the materials under study. A precise characterisation of the anisotropy of quartz would require a more detailed modeling using matrix formalisms to account for the transmission through the quartz layer and Jones calculus. In any case, the magnetic anisotropy controlled spintronic emitter opens as such an effective way to implement time domain ellipsometry at THz frequencies. This technique has during the last decade established itself as a powerful THz spectroscopic tool allowing to bypass the need for a measurement against a standard reference sample, having a larger spectral coverage than frequency domain spectroscopy, and providing direct access to phase information\cite{Agulto2021,Nagashima2013}. Of the remaining challenges to be solved, the quality of the THz polarizing optical component needed to obtain the ellipsometric functions, is an important one\cite{Neshat2013}. Using the here proposed emitter, a source with easily variable s- and p-polarization becomes available that is moreover intrinsically very broadband. As such the polarization analysis in a time-domain spintronic THz ellipsometer would not require a rotating analyser anymore as all polarization states can easily be generated upon emission.


\section{Conclusion}
\label{sec:conclusion}

In this work we have demonstrated a conceptually new approach for polarization control of a high-performance broadband terahertz source based on spintronic generation of ultrafast current dipoles. Our approach combines the intrinsic polarization versatility of inverse spin Hall terahertz emitters (i.e. emitted polarization strictly orthogonal to their magnetization state) with scalar field Stoner-Wohlfarth magnetization rotation offered by an uniaxially anisotropic ferromagnetic spin pumping layer. To achieve this, a rare-earth based intermetallic thin film heterostructure has been developed. Exchange coupling between a strongly anisotropic hard magnetic TbCo$_2$ thin film and soft magnetic isotropic CoFe layers leads to an effective ferromagnetic heterostructure with a lowered anisotropy field (down to 5kA/m) with an average saturation magnetization (close to 60\% of the CoFe alloy). Sandwiched between W and Pt layers as spinorbit coupling metals, this anisotropic spin pumping layer achieves THz strengths comparable to the strongest STEs reported with similar Fourier-limited bandwidths while simultaneously allowing polarization rotation over a full 2$\pi$ radians with just a uniaxial field sweep of 5kA/m (or $\mu_0 H = 6.5$mT). A practical extended model for the Stoner-Wohlfarth mediated polarization rotation allows to set on-demand any linear THz polarization by applying the appropriate hard axis field. As such this anisotropic spintronic terahertz emitter can act as a new versatile source for terahertz time domain broadband ellipsometry. This has been qualitatively demonstrated on a quartz sample. But the impact of the easy polarization control of this emitter type has several other far reaching applications. In this work the Stoner-Wohlfarth type magnetization control was mainly quasistatically exploited. Dynamic fast control of the reversible magnetization is imaginable by operating the uniaxial anisotropic emitter near the spin reorientation transition (SRT), where the magnetization along the easy axis moves unhindered by an energy barrier under small amplitude RF excitations. Modulation speeds well beyond 10MHz are expected \cite{TiercelinPreobrazhensky2009, KlimovIgnatov2010}. Such a wideband fast-modulated STE with near 100\% modulation index is a missing component for wireless THz datacom but could also find use in low-noise modulation ellipsometric THz spectroscopy. Besides implementing SRT modulation in these anisotropic rare-earth based intermetallic spintronic emitters, a further perspective is offered by the magnetoelastic properties of such systems. These structures have been demonstrated to exhibit giant magnetostriction \cite{DebrayLudwig2004}. As a result the anisotropic rotation of the THz polarization could be controlled by magnetoelectric coupling of the stress in the rare-earth based emitters with a piezoelectric substrate. As such, this would completely remove the need for magnetic biasing to rotate the polarization, which could now be achieved by proper voltage gating of a piezoelectric carrier.
In any case, it is clear that within the research activity on terahertz spintronics, the use of engineered magnetic anisotropic emitters is a fertile playing ground with a perspective towards several improved terahertz functionalities.




\begin{backmatter}
\bmsection{Funding}
This work was funded by FET project s-NEBULA/Grant No. 863155. Authors acknowledge doctoral grant competition
CZ.02.2.69/0.0/0.0/19\_073/0016945 under project
DGS/TEAM/2020-027.

\bmsection{Acknowledgments}
The authors thank the RENATECH Network for the support in the realization of the devices.

\bmsection{Disclosures}
The authors declare no conflicts of interest.





\bmsection{Data Availability Statement}
The measurement data are publicly available through the ZENODO server under the reference DOI:~10.5072/zenodo.957553

\bmsection{Supplemental document}
See Supplement for supporting content. 

\end{backmatter}

\bibliography{bibliography.bib,pierre.bib}

\begin{thebibliography}{10}
\newcommand{\enquote}[1]{``#1''}

\bibitem{Dhillon2017}
S.~S. Dhillon, M.~S. Vitiello, E.~H. Linfield, A.~G. Davies, M.~C. Hoffmann,
  J.~Booske, C.~Paoloni, M.~Gensch, P.~Weightman, G.~P. Williams,
  E.~Castro-Camus, D.~R.~S. Cumming, F.~Simoens, I.~Escorcia-Carranza,
  J.~Grant, S.~Lucyszyn, M.~Kuwata-Gonokami, K.~Konishi, M.~Koch, C.~A.
  Schmuttenmaer, T.~L. Cocker, R.~Huber, A.~G. Markelz, Z.~D. Taylor, V.~P.
  Wallace, J.~A. Zeitler, J.~Sibik, T.~M. Korter, B.~Ellison, S.~Rea,
  P.~Goldsmith, K.~B. Cooper, R.~Appleby, D.~Pardo, P.~G. Huggard, V.~Krozer,
  H.~Shams, M.~Fice, C.~Renaud, A.~Seeds, A.~Stöhr, M.~Naftaly, N.~Ridler,
  R.~Clarke, J.~E. Cunningham, and M.~B. Johnston, \enquote{The 2017 terahertz
  science and technology roadmap,} {\protect\JournalTitle{Journal of Physics D:
  Applied Physics}} \textbf{50}, 043001 (2017).

\bibitem{Hindle2018}
F.~Hindle, L.~Kuuliala, M.~Mouelhi, A.~Cuisset, C.~Bray, M.~Vanwolleghem,
  F.~Devlieghere, G.~Mouret, and R.~Bocquet, \enquote{{Monitoring of food
  spoilage by high resolution THz analysis},} {\protect\JournalTitle{Analyst}}
  \textbf{143}, 5536--5544 (2018).

\bibitem{Jepsen2011}
P.~U. Jepsen, D.~G. Cooke, and M.~Koch, \enquote{{Terahertz spectroscopy and
  imaging - Modern techniques and applications},} {\protect\JournalTitle{Laser
  and Photonics Reviews}} \textbf{5}, 124--166 (2011).

\bibitem{Ellrich2020}
F.~Ellrich, .~M. Bauer, .~N. Schreiner, .~A. Keil, T.~Pfeiffer, J.~Klier, .~S.
  Weber, J.~Jonuscheit, F.~Friederich, and .~D. Molter, \enquote{Terahertz
  quality inspection for automotive and aviation industries,}
  {\protect\JournalTitle{Journal of Infrared, Millimeter, and Terahertz Waves}}
  \textbf{41}, 470--489 (2020).

\bibitem{nagatsuma2016advances}
T.~Nagatsuma, G.~Ducournau, and C.~C. Renaud, \enquote{Advances in terahertz
  communications accelerated by photonics,} {\protect\JournalTitle{Nature
  Photonics}} \textbf{10}, 371--379 (2016).

\bibitem{Shan2009}
J.~Shan, J.~I. Dadap, and T.~F. Heinz, \enquote{{Circularly polarized light in
  the single-cycle limit: The nature of highly polychromatic radiation of
  defined polarization},} {\protect\JournalTitle{Optics Express}} \textbf{17},
  7431 (2009).

\bibitem{Konishi2020}
K.~Konishi, T.~Kan, and M.~Kuwata-Gonokami, \enquote{{Tunable and nonlinear
  metamaterials for controlling circular polarization},}
  {\protect\JournalTitle{Journal of Applied Physics}} \textbf{127} (2020).

\bibitem{Grady2013}
N.~K. Grady, J.~E. Heyes, D.~R. Chowdhury, Y.~Zeng, M.~T. Reiten, A.~K. Azad,
  A.~J. Taylor, D.~A. Dalvit, and H.~T. Chen, \enquote{{Terahertz metamaterials
  for linear polarization conversion and anomalous refraction},}
  {\protect\JournalTitle{Science}} \textbf{340}, 1304--1307 (2013).

\bibitem{Markovich2013}
D.~L. Markovich, A.~Andryieuski, M.~Zalkovskij, R.~Malureanu, and A.~V.
  Lavrinenko, \enquote{{Metamaterial polarization converter analysis: Limits of
  performance},} {\protect\JournalTitle{Applied Physics B: Lasers and Optics}}
  \textbf{112}, 143--152 (2013).

\bibitem{Amer2005}
N.~Amer, W.~C. Hurlbut, B.~J. Norton, Y.~S. Lee, and T.~B. Norris,
  \enquote{{Generation of terahertz pulses with arbitrary elliptical
  polarization},} {\protect\JournalTitle{Applied Physics Letters}} \textbf{87},
  1--3 (2005).

\bibitem{Shimano2005}
R.~Shimano, H.~Nishimura, and T.~Sato, \enquote{{Frequency tunable circular
  polarization control of Terahertz radiation},}
  {\protect\JournalTitle{Japanese Journal of Applied Physics, Part 2: Letters}}
  \textbf{44}, 19--22 (2005).

\bibitem{Lee2012}
K.~Lee, M.~Yi, J.~D. Song, and J.~Ahn, \enquote{{Polarization shaping of
  few-cycle terahertz waves},} {\protect\JournalTitle{Optics Express}}
  \textbf{20}, 12463 (2012).

\bibitem{Sato2013}
M.~Sato, T.~Higuchi, N.~Kanda, K.~Konishi, K.~Yoshioka, T.~Suzuki, K.~Misawa,
  and M.~Kuwata-Gonokami, \enquote{{Terahertz polarization pulse shaping with
  arbitrary field control},} {\protect\JournalTitle{Nature Photonics}}
  \textbf{7}, 724--731 (2013).

\bibitem{Hirota2005}
Y.~Hirota, M.~Tani, and M.~Hangyo, \enquote{{Polarization modulation of
  terahertz wave by multipole type photoconductive antenna},}
  {\protect\JournalTitle{IQEC, International Quantum Electronics Conference
  Proceedings}} \textbf{2005}, 1767--1768 (2005).

\bibitem{Dai2009}
J.~Dai, N.~Karpowicz, and X.~C. Zhang, \enquote{{Coherent polarization control
  of terahertz waves generated from two-color laser-induced gas plasma},}
  {\protect\JournalTitle{Physical Review Letters}} \textbf{103}, 1--4 (2009).

\bibitem{Wen2009}
H.~Wen and A.~M. Lindenberg, \enquote{Coherent terahertz polarization control
  through manipulation of electron trajectories,}
  {\protect\JournalTitle{Physical Review Letters}} \textbf{103}, 2--5 (2009).

\bibitem{You2013}
Y.~S. You, T.~I. Oh, and K.-Y. Kim, \enquote{{Mechanism of elliptically
  polarized terahertz generation in two-color laser filamentation},}
  {\protect\JournalTitle{Optics Letters}} \textbf{38}, 1034--1036 (2013).

\bibitem{seifert2016}
T.~Seifert, S.~Jaiswal, U.~Martens, J.~Hannegan, L.~Braun, P.~Maldonado,
  F.~Freimuth, A.~Kronenberg, J.~Henrizi, I.~Radu \emph{et~al.},
  \enquote{Efficient metallic spintronic emitters of ultrabroadband terahertz
  radiation,} {\protect\JournalTitle{Nat. Photonics}} \textbf{10}, 483--488
  (2016).

\bibitem{Yang2016}
D.~Yang, J.~Liang, C.~Zhou, L.~Sun, R.~Zheng, S.~Luo, Y.~Wu, and J.~Qi,
  \enquote{{Powerful and tunable THz emitters based on the Fe/Pt magnetic
  heterostructure},} {\protect\JournalTitle{Advanced Optical Materials}}
  \textbf{4}, 1944--1949 (2016).

\bibitem{Seifert2017}
T.~Seifert, S.~Jaiswal, M.~Sajadi, G.~Jakob, S.~Winnerl, M.~Wolf,
  M.~Kl{\"{a}}ui, and T.~Kampfrath, \enquote{{Ultrabroadband single-cycle
  terahertz pulses with peak fields of 300 kV cm-1 from a metallic spintronic
  emitter},} {\protect\JournalTitle{Applied Physics Letters}} \textbf{110}
  (2017).

\bibitem{Fulop2020}
J.~A. F{\"{u}}l{\"{o}}p, S.~Tzortzakis, and T.~Kampfrath, \enquote{Laser-driven
  strong-field terahertz sources,} {\protect\JournalTitle{Advanced Optical
  Materials}} \textbf{8}, 1--25 (2020).

\bibitem{Gueckstock2021}
O.~Gueckstock, L.~Nádvorník, T.~Seifert, M.~Borchert, G.~Jakob, G.~Schmidt,
  G.~Woltersdorf, M.~Kläui, M.~Wolf, and T.~Kampfrath, \enquote{{Modulating
  the polarization of broadband terahertz pulses from a spintronic emitter at
  rates up to 10 kHz},} {\protect\JournalTitle{Optica}} \textbf{8} (2021).

\bibitem{KongAOM2019}
D.~Kong, X.~Wu, B.~B. Wang, T.~Nie, M.~Xiao, C.~Pandey, Y.~Gao, L.~Wen,
  W.~Zhao, C.~Ruan, J.~Miao, Y.~Li, and L.~Wang, \enquote{Broadband spintronic
  terahertz emitter with magnetic-field manipulated polarizations,}
  {\protect\JournalTitle{Advanced Optical Materials}} \textbf{7}, 1--9 (2019).

\bibitem{Hibberd2019}
M.~T. Hibberd, D.~S. Lake, N.~A. Johansson, T.~Thomson, S.~P. Jamison, and
  D.~M. Graham, \enquote{{Magnetic-field tailoring of the terahertz
  polarization emitted from a spintronic source},}
  {\protect\JournalTitle{Applied Physics Letters}} \textbf{114} (2019).

\bibitem{ChenAPL2019}
X.~Chen, X.~Wu, S.~Shan, F.~Guo, D.~Kong, C.~Wang, T.~Nie, C.~Pandey, L.~Wen,
  W.~Zhao, C.~Ruan, J.~Miao, Y.~Li, and L.~Wang, \enquote{{Generation and
  manipulation of chiral broadband terahertz waves from cascade spintronic
  terahertz emitters},} {\protect\JournalTitle{Applied Physics Letters}}
  \textbf{115} (2019).

\bibitem{Stoner1948}
E.~C. Stoner and E.~P. Wohlfarth, \enquote{{A mechanism of magnetic hysteresis
  in heterogeneous alloys},} {\protect\JournalTitle{Philosophical Transactions
  of the Royal Society of London. Series A, Mathematical and Physical
  Sciences}} \textbf{240}, 599--642 (1948).

\bibitem{Kampfrathnatphot2013}
T.~Kampfrath, K.~Tanaka, and K.~A. Nelson, \enquote{{Resonant and nonresonant
  control over matter and light by intense terahertz transients},}
  {\protect\JournalTitle{Nature Photonics}} \textbf{7}, 680--690 (2013).

\bibitem{Cocker2013}
T.~L. Cocker and R.~Huber, \enquote{{Ultrafast pulse shaping: A new twist on
  terahertz pulses},} {\protect\JournalTitle{Nature Photonics}} \textbf{7},
  678--679 (2013).

\bibitem{Mosley2017}
C.~D. Mosley, M.~Failla, D.~Prabhakaran, and J.~Lloyd-Hughes,
  \enquote{{Terahertz spectroscopy of anisotropic materials using beams with
  rotatable polarization},} {\protect\JournalTitle{Scientific Reports}}
  \textbf{7}, 1--9 (2017).

\bibitem{LeGall2000}
H.~{Le Gall}, J.~{Ben Youssef}, F.~Socha, N.~Tiercelin, V.~Preobrazhensky, and
  P.~Pernod, \enquote{{Low field anisotropic magnetostriction of single domain
  exchange-coupled (TbFe/Fe) multilayers},} {\protect\JournalTitle{Journal of
  Applied Physics}} \textbf{87}, 5783--5785 (2000).

\bibitem{BenYoussef2002}
J.~B. Youssef, N.~Tiercelin, F.~Petit, H.~{Le Gall}, V.~Preobrazhensky, and
  P.~Pernod, \enquote{{Statics and dynamics in giant magnetostrictive $Tb_x
  Fe_{1-x}/Fe_{0.6}Co_{0.4}$ multilayers for MEMS},}
  {\protect\JournalTitle{IEEE Transactions on Magnetics}} \textbf{38},
  2817--2819 (2002).

\bibitem{Hoffmann2013}
A.~Hoffmann, \enquote{{Spin hall effects in metals},}
  {\protect\JournalTitle{IEEE Transactions on Magnetics}}  (2013).

\bibitem{carli1977reflectivity}
B.~Carli, \enquote{Reflectivity of metallic films in the infrared,}
  {\protect\JournalTitle{JOSA}} \textbf{67}, 908--910 (1977).

\bibitem{kroll2007metallic}
J.~Kr{\"o}ll, J.~Darmo, and K.~Unterrainer, \enquote{Metallic wave-impedance
  matching layers for broadband terahertz optical systems,}
  {\protect\JournalTitle{Opt. Express}} \textbf{15}, 6552--6560 (2007).

\bibitem{neumann2016temperature}
L.~Neumann, D.~Meier, J.~Schmalhorst, K.~Rott, G.~Reiss, and M.~Meinert,
  \enquote{{Temperature dependence of the spin Hall angle and switching current
  in the nc-W (O)/CoFeB/MgO system with perpendicular magnetic anisotropy},}
  {\protect\JournalTitle{Applied Physics Letters}} \textbf{109}, 142405 (2016).

\bibitem{hao2015giant}
Q.~Hao and G.~Xiao, \enquote{{Giant spin Hall effect and switching induced by
  spin-transfer torque in a W/Co 40 Fe 40 B 20/MgO structure with perpendicular
  magnetic anisotropy},} {\protect\JournalTitle{Physical Review Applied}}
  \textbf{3}, 034009 (2015).

\bibitem{explanation_tilt}
Perfect monodomain rotation is an idealized picture that is in principle only
  attained at perfect orthogonality between the applied field and the
  anisotropy axis. In practice, the anisotropy axis is intentionally slightly
  tilted to impose a rotation sense during a hysteresis cycle and avoid
  collapse of the latter. In our experiments the easy axis was intentionally
  tilted by $\approx 10^\circ$. This causes an abrupt magnetization flip at a
  critical field $H_c$ lower than the anisotropy field $H_A$. This flip is
  observed both in the behaviour of the polarization azimuth and the field
  rotation.

\bibitem{wiregrid}
{PureWavePolarizers}, \enquote{5 micron wire grid polarizer,}
  https://www.purewavepolarizers.com/
  wire-grid-polarizers/5-micron-wire-far-ir-thz-polarizer (2021).

\bibitem{KlimovIgnatov2010}
A.~Klimov, Y.~Ignatov, N.~Tiercelin, V.~Preobrazhensky, P.~Pernod, and
  S.~Nikitov, \enquote{Ferromagnetic resonance and magnetoelastic demodulation
  in thin active films with an uniaxial anisotropy,} {\protect\JournalTitle{J.
  Appl. Phys.}} \textbf{107}, 093916--6 (2010).

\bibitem{Dang2020}
T.~H. Dang, J.~Hawecker, E.~Rongione, G.~Baez~Flores, D.~Q. To, J.~C.
  Rojas-Sanchez, H.~Nong, J.~Mangeney, J.~Tignon, F.~Godel, S.~Collin,
  P.~Seneor, M.~Bibes, A.~Fert, M.~Anane, J.-M. George, L.~Vila,
  M.~Cosset-Cheneau, D.~Dolfi, R.~Lebrun, P.~Bortolotti, K.~Belashchenko,
  S.~Dhillon, and H.~Jaffrès, \enquote{Ultrafast spin-currents and charge
  conversion at 3d-5d interfaces probed by time-domain terahertz spectroscopy,}
  {\protect\JournalTitle{Applied Physics Reviews}} \textbf{7}, 041409 (2020).

\bibitem{torosyan2018}
G.~Torosyan, S.~Keller, L.~Scheuer, R.~Beigang, and E.~T. Papaioannou,
  \enquote{{Optimized spintronic terahertz emitters based on epitaxial grown
  Fe/Pt layer structures},} {\protect\JournalTitle{Sci. Rep.}} \textbf{8}, 1311
  (2018).

\bibitem{cullity2011introduction}
B.~D. Cullity and C.~D. Graham, \emph{Introduction to magnetic materials} (John
  Wiley \& Sons, 2011).

\bibitem{vaz2008magnetism}
C.~Vaz, J.~Bland, and G.~Lauhoff, \enquote{Magnetism in ultrathin film
  structures,} {\protect\JournalTitle{Reports on Progress in Physics}}
  \textbf{71}, 056501 (2008).

\bibitem{arrott1991phenomenology}
A.~S. Arrott and B.~Heinrich, \enquote{Phenomenology of anisotropy in the
  ferromagnetism of ultrathin films,} {\protect\JournalTitle{Journal of
  magnetism and magnetic materials}} \textbf{93}, 571--586 (1991).

\bibitem{kools1996exchange}
J.~Kools, \enquote{Exchange-biased spin-valves for magnetic storage,}
  {\protect\JournalTitle{IEEE transactions on magnetics}} \textbf{32},
  3165--3184 (1996).

\bibitem{boyd1960magnetic}
E.~L. Boyd, \enquote{Magnetic anisotropy in single-crystal thin films,}
  {\protect\JournalTitle{IBM Journal of Research and Development}} \textbf{4},
  116--129 (1960).

\bibitem{heinrich2006ultrathin}
B.~Heinrich and J.~A.~C. Bland, \emph{Ultrathin magnetic structures II:
  Measurement techniques and novel magnetic properties}, vol.~2 (Springer
  Science \& Business Media, 2006).

\bibitem{Grischkowsky1990}
D.~Grischkowsky, S.~Keiding, M.~van Exter, and C.~Fattinger,
  \enquote{Far-infrared time-domain spectroscopy with terahertz beams of
  dielectrics and semiconductors,} {\protect\JournalTitle{Journal of the
  Optical Society of America B}} \textbf{7}, 2006 (1990).

\bibitem{Davies2018}
C.~L. Davies, J.~B. Patel, C.~Q. Xia, L.~M. Herz, and M.~B. Johnston,
  \enquote{Temperature-dependent refractive index of quartz at terahertz
  frequencies,} {\protect\JournalTitle{Journal of Infrared, Millimeter, and
  Terahertz Waves}} \textbf{39}, 1236--1248 (2018).

\bibitem{planken2001}
P.~C. Planken, H.-K. Nienhuys, H.~J. Bakker, and T.~Wenckebach,
  \enquote{{Measurement and calculation of the orientation dependence of
  terahertz pulse detection in ZnTe},} {\protect\JournalTitle{JOSA B}}
  \textbf{18}, 313--317 (2001).

\bibitem{castro2009extraction}
E.~Castro-Camus and M.~Johnston, \enquote{Extraction of the anisotropic
  dielectric properties of materials from polarization-resolved terahertz
  time-domain spectra,} {\protect\JournalTitle{Journal of Optics A: Pure and
  Applied Optics}} \textbf{11}, 105206 (2009).

\bibitem{Agulto2021}
V.~C. Agulto, T.~Iwamoto, H.~Kitahara, K.~Toya, V.~K. Mag-usara, M.~Imanishi,
  Y.~Mori, M.~Yoshimura, and M.~Nakajima, \enquote{{Terahertz time-domain
  ellipsometry with high precision for the evaluation of GaN crystals with
  carrier densities up to 10$^{20}$cm$^{-3}$},}
  {\protect\JournalTitle{Scientific Reports}} \textbf{11}, 18129 (2021).

\bibitem{Nagashima2013}
T.~Nagashima, M.~Tani, and M.~Hangyo, \enquote{{Polarization-sensitive THz-TDS
  and its Application to Anisotropy Sensing},} {\protect\JournalTitle{Journal
  of Infrared, Millimeter, and Terahertz Waves}} \textbf{34}, 740--775 (2013).

\bibitem{Neshat2013}
M.~Neshat and N.~P. Armitage, \enquote{{Developments in THz range
  ellipsometry},} {\protect\JournalTitle{Journal of Infrared, Millimeter, and
  Terahertz Waves}} \textbf{34}, 682--708 (2013).

\bibitem{TiercelinPreobrazhensky2009}
N.~Tiercelin, V.~Preobrazhensky, V.~Mortet, A.~Talbi, A.~Soltani, K.~Haenen,
  and P.~Pernod, \enquote{Thin film magnetoelectric composites near spin
  reorientation transition,} {\protect\JournalTitle{Journal of Magnetism and
  Magnetic Materials}} \textbf{321}, 1803--1807 (2009).

\bibitem{DebrayLudwig2004}
A.~Debray, A.~Ludwig, T.~Bourouina, A.~Asaoka, N.~Tiercelin, G.~Reyne, T.~Oki,
  E.~Quandt, H.~Muro, and H.~Fujita, \enquote{Application of a multilayered
  magnetostrictive film to a micromachined 2-d optical scanner,}
  {\protect\JournalTitle{Microelectromechanical Systems, Journal of}}
  \textbf{13}, 264--271 (2004).

\end{thebibliography}






\end{document}